# Hall Shift Current and Nonlinear Anomalous Hall Effect in Gapped Dirac Fermion Systems

Toshihito Osada*

*Department of Applied Physics, The University of Tokyo,*

*7-3-1 Hongo, Bunkyo-ku, Tokyo 113-8656, Japan.*

We discuss a new mechanism for the nonlinear anomalous Hall effect in tilted two-dimensional gapped Dirac fermion systems. This mechanism originates from a side-jump displacement associated with Landau-Zener tunneling across the mass gap. In lightly doped tilted Dirac systems with a small gap, this interband mechanism can contribute to the nonlinear anomalous Hall effect in addition to the conventional intraband mechanism arising from the Berry curvature dipole. It provides a possible explanation for the nonlinear Hall effect accompanied by nonlinear longitudinal transport observed in the organic Dirac fermion system $\alpha$-$(ET)_2I_3$ with an extremely small gap.

The nonlinear anomalous Hall effect (NLAHE), or nonlinear Hall effect, is a topological transport phenomenon in which a Hall current proportional to the square of the electric field appears at zero magnetic field in nonmagnetic crystals with time-reversal symmetry (TRS). The intrinsic anomalous Hall effect (AHE) is caused by the summation of the anomalous velocity originating from the Berry curvature over the occupied states. It vanishes in systems with TRS because the Berry curvature contributions from the occupied states cancel. In a current-carrying state under an electric field, however, the distribution function is modified and the balance of Berry curvature contributions is broken, resulting in the NLAHE. This is the conventional intraband mechanism of the NLAHE arising from the Berry curvature dipole (BCD) [1]. So far, the NLAHE has been observed in various topological metallic systems [2].

The NLAHE was also observed in the organic Dirac fermion system $\alpha$-$(ET)_2I_3$, where ET denotes BEDT-TTF [3]. $\alpha$-$(ET)_2I_3$ is a layered organic conductor with very weak interlayer coupling, less than 1 meV, and is therefore usually regarded as a two-dimensional (2D) system. At ambient pressure, it undergoes a transition to a charge-ordered (CO) insulator phase at 135 K. The CO phase is suppressed by applying pressure and disappears above the critical pressure of approximately 1.2 GPa, where the system becomes a 2D massless Dirac fermion system with a pair of tilted Dirac cones [4]. The NLAHE was observed in the weak CO state just below the critical pressure [3]. This weak CO state was experimentally clarified to be a 2D massive Dirac fermion system with a small CO gap at the Dirac point [5]. It differs from other NLAHE systems in that it has an extremely small gap of approximately 0.2 meV and behaves as a weak insulator doped with a small number of carriers, typically of the order of $10^{17}$ $cm^{-3}$.

In this system, in addition to the conventional intraband mechanism, an interband mechanism associated with Landau-Zener (LZ) tunneling [6, 7] across the gap may also operate. In a situation where the NLAHE arises only from the intraband mechanism, the longitudinal conduction is expected to remain Ohmic. In $\alpha$-$(ET)_2I_3$, however, preliminary measurements indicate that nonlinear longitudinal conduction also appears in the same regime, suggesting a contribution from tunneling conduction [8].

The interband mechanism discussed here is based on the theory of nonreciprocal transport in a one-dimensional (1D) LZ tunneling system developed by Kitamura et al. [9, 10]. They introduced a shift-vector correction to interband tunneling in a 1D two-band system without space-inversion symmetry (SIS). The shift vector has already been used to explain the light-induced shift current in the bulk photovoltaic effect [11, 12].

The scenario of the present interband mechanism is as follows. When an electron wave packet in the valence band v is driven in the positive $k_x$ direction by an electric field $\mathbf{E}$ applied along the negative $x$ direction, following the equation of motion, $\hbar d\mathbf{k}/dt = (-e)\mathbf{E}$, the electron is transferred to the conduction band c through LZ tunneling with a finite transition rate $dP(\mathbf{k})/dt = \{\partial P(\mathbf{k})/\partial \mathbf{k}\}\cdot(d\mathbf{k}/dt)$, where $P(\mathbf{k})$ denotes the transient occupation probability of the conduction band at $\mathbf{k}$. In real space, the wave packet splits into two packets in the conduction and valence bands. Their center coordinates differ within a unit cell by the shift vector

$$\mathbf{R}_{\mathrm{cv}}^{(x)}(\mathbf{k}) = \mathbf{A}_{\mathrm{cc}}(\mathbf{k}) - \mathbf{A}_{\mathrm{vv}}(\mathbf{k}) - \frac{\partial}{\partial \mathbf{k}} \arg[\mathbf{A}_{\mathrm{cv}}(\mathbf{k})]_x , \tag{1}$$

where $\mathbf{A}_{mn}(\mathbf{k}) = i\langle u_m(\mathbf{k})|(\partial/\partial\mathbf{k})u_n(\mathbf{k})\rangle$ is the Berry connection, and $|u_n(\mathbf{k})\rangle$ is the periodic part of the Bloch state in band $n$. $[\mathbf{A}]_\alpha$ denotes the $\alpha$-component of the vector $\mathbf{A}$. The schematic illustration of the shift vector in LZ tunneling is shown in Fig. 1(a). The $y$- component of the shift vector $[\mathbf{R}_{\mathrm{cv}}{}^{(x)}(\mathbf{k})]_y$ gives the side-jump of the electron in the Hall direction perpendicular to the driving $x$ direction. If the sum of $[\mathbf{R}_{\mathrm{cv}}{}^{(x)}(\mathbf{k})]_y dP(\mathbf{k})/dt$ over the $\mathbf{k}$ states where transitions are allowed is finite, a Hall shift current appears, resulting in a contribution to the NLAHE.

We consider the following model of a 2D Dirac fermion system,

$$H_{\mathrm{R}}(\mathbf{k}) = d_0(k_x)\sigma_0 + \mathbf{d}(\mathbf{k})\cdot\boldsymbol{\sigma} = \hbar v_0 k_x \sigma_0 + \hbar v_x k_x \sigma_x + \hbar v_y k_y \sigma_y + \Delta\sigma_z, \qquad (2)$$

where Pauli matrices $\boldsymbol{\sigma} = (\sigma_x, \sigma_y, \sigma_z)$ represent the sublattice pseudospin, and $\mathbf{d}(\mathbf{k}) = |\mathbf{d}(\mathbf{k})|\mathbf{n}_{\mathbf{d}(\mathbf{k})} = (\hbar v_x k_x, \hbar v_y k_y, \Delta)$. We assume $\Delta > 0$ without losing any generality. The dispersions of the conduction band $E_{\mathrm{c}}(\mathbf{k})$ and the valence band $E_{\mathrm{v}}(\mathbf{k})$ are given by

$$E_{c,v}(\mathbf{k}) = d_0(k_x) \pm |\mathbf{d}(\mathbf{k})| = \hbar v_0 k_x \pm \sqrt{\Delta^2 + (\hbar v_x k_x)^2 + (\hbar v_y k_y)^2}\,. \qquad (3)$$

They correspond to Dirac cones with the tilt ratio $\eta = v_0/v_x$ and an indirect energy gap of $2(1-\eta^2)^{1/2}\Delta$ between the band extrema for $|\eta| < 1$. The direct gap at the Dirac point is $2\Delta$. $H_{\mathrm{R}}(\mathbf{k})$ represents one valley with a right-handed chirality. In a TRS system, there exists a time-reversal valley, which is written in the same bases as

$$H_{\mathrm{L}}(\mathbf{k}) = H_{\mathrm{R}}{}^*(-\mathbf{k}) = -\hbar v_0 k_x \sigma_0 - \hbar v_x k_x \sigma_x + \hbar v_y k_y \sigma_y + \Delta\sigma_z\,. \qquad (4)$$

We consider the situation in which, under the electric field in the negative $x$ direction, electrons in the valence band are adiabatically driven in the positive $k_x$ direction, and undergo Landau-Zener transitions to the conduction band with finite probability. In a two-level system $H(\mathbf{k}) = \mathbf{d}(\mathbf{k})\cdot\boldsymbol{\sigma}$, the norm of the $\alpha$ component of the interband Berry connection and the $\beta$ component of the shift vector for $\alpha$-direction driving are generally given by [9]

$$|[\mathbf{A}_{\mathrm{cv}}(\mathbf{k})]_\alpha| = \left|\mathbf{d}(\mathbf{k}) \times \frac{\partial \mathbf{d}(\mathbf{k})}{\partial k_\alpha}\right| \Big/ \left\{2|\mathbf{d}(\mathbf{k})|^2\right\}, \qquad (5a)$$

$$[\mathbf{R}_{\mathrm{cv}}^{(\alpha)}(\mathbf{k})]_\beta = \left\{ \left( \mathbf{n}_{\mathbf{d}(\mathbf{k})} \times \frac{\partial \mathbf{n}_{\mathbf{d}(\mathbf{k})}}{\partial k_\alpha} \right) \cdot \frac{\partial^2 \mathbf{n}_{\mathbf{d}(\mathbf{k})}}{\partial k_\alpha \partial k_\beta} \right\} \Big/ \left| \mathbf{n}_{\mathbf{d}(\mathbf{k})} \times \frac{\partial \mathbf{n}_{\mathbf{d}(\mathbf{k})}}{\partial k_\alpha} \right|^2 . \quad (5b)$$

Here, $\mathbf{n}_{\mathbf{d}(\mathbf{k})}$ denotes the unit vector along $\mathbf{d}(\mathbf{k})$. For the present Dirac system, we obtain

$$\left| [\mathbf{A}_{\mathrm{cv}}(\mathbf{k})]_x \right| = \frac{\hbar v_x \sqrt{\Delta^2 + (\hbar v_y k_y)^2}}{2\left\{ \Delta^2 + (\hbar v_x k_x)^2 + (\hbar v_y k_y)^2 \right\}} , \quad (6a)$$

$$\begin{cases} [\mathbf{R}_{\mathrm{cv}}^{(x)}(\mathbf{k})]_x = 0 \\ [\mathbf{R}_{\mathrm{cv}}^{(x)}(\mathbf{k})]_y = -\dfrac{\hbar v_y \Delta (\hbar v_x k_x)}{\left\{ \Delta^2 + (\hbar v_y k_y)^2 \right\} \sqrt{\Delta^2 + (\hbar v_x k_x)^2 + (\hbar v_y k_y)^2}} \end{cases} . \quad (6b)$$

Note that the shift vector $\mathbf{R}_{\mathrm{cv}}^{(x)}(\mathbf{k})$ is always perpendicular to the driving $x$ direction. The $y$ component $[\mathbf{R}_{\mathrm{cv}}^{(x)}(\mathbf{k})]_y$ gives the transverse side jump in the Hall direction associated with the LZ transition.

The LZ transition process of a single electron is represented by the time-dependent Schrödinger equation. When $\mathbf{k}$ is adiabatically driven by the force $(-e)\mathbf{E}$ in the $x$ direction, the wave function $u(\mathbf{k})$ is written as a linear combination of the conduction- and valence-band states,

$$\left| u(\mathbf{k}) \right\rangle = a_c(\mathbf{k}) e^{i\Gamma_c(\mathbf{k})} \left| u_c(\mathbf{k}) \right\rangle + a_v(\mathbf{k}) e^{i\Gamma_v(\mathbf{k})} \left| u_v(\mathbf{k}) \right\rangle , \quad (7)$$

$$\Gamma_n(\mathbf{k}) \equiv -\int_{k_0}^{k_x} \frac{E_n(k_x', k_y)}{(-e)E_x} dk_x' + \int_{k_0}^{k_x} [\mathbf{A}_{nn}(k', k_y)]_x dk_x' . \quad (8)$$

Substitution of this expression into the time-dependent Schrödinger equation leads to

$$\begin{aligned} \frac{\partial a_c(\mathbf{k})}{\partial k_x} &= i \left| [\mathbf{A}_{cv}(\mathbf{k})]_x \right| e^{i\Theta_{cv}(\mathbf{k})} a_v(\mathbf{k}) \\ \frac{\partial a_v(\mathbf{k})}{\partial k_x} &= i \left| [\mathbf{A}_{cv}(\mathbf{k})]_x \right| e^{-i\Theta_{cv}(\mathbf{k})} a_c(\mathbf{k}) \\ \frac{\partial \Theta_{cv}(\mathbf{k})}{\partial k_x} &= \frac{E_c(\mathbf{k}) - E_v(\mathbf{k})}{(-e)E_x} - [\mathbf{R}_{cv}^{(x)}(\mathbf{k})]_x \end{aligned} \quad (9)$$

Using Eq. (6), we numerically solved the above equation with the initial conditions $a_c(k_x \rightarrow -\infty, k_y) = 0$ and $a_v(k_x \rightarrow -\infty, k_y) = 1$. The calculated conduction-band occupation probability $P(\mathbf{k}) = |a_c(\mathbf{k})|^2$ is plotted as a function of $k_x$ at fixed $k_y$ for several electric field values by the blue solid curves in Fig. 1(b), and as a function of $\mathbf{k}$ for one electric field in Fig.1(c). $P(\mathbf{k})$ increases from zero to a finite value with increasing $k_x$ around the Dirac point $\mathbf{k} = 0$. In the limit $k_x \rightarrow \infty$, $P(\mathbf{k})$ converges to the LZ probability [7],

$$P_{\mathrm{LZ}}(k_y) = \exp\left[-\frac{\pi\left\{\Delta^2 + (\hbar v_y k_y)^2\right\}}{\hbar v_x e|E_x|}\right]. \tag{10}$$

In the 2D Dirac system, the LZ probability is not affected by the shift vector because $[\mathbf{R}_{cv}^{(x)}(\mathbf{k})]_x = 0$, in contrast to the 1D case discussed by Kitamura et al. We can see transient oscillations of $P(\mathbf{k})$ at small electric fields, corresponding to coherent oscillations in single-passage LZ dynamics. They originate from beating between the conduction- and valence-band states in the time domain [13, 14].

When electrons are scattered in the conduction and valence bands and lose their coherence, these interference oscillations are smeared out. For the incoherent case, which is relevant to dissipative steady-state transport, we use a phenomenological non-oscillatory probability $P'(\mathbf{k})$ instead of the coherent single-passage probability $P(\mathbf{k})$. We employ $P'(\mathbf{k})$ with the same $\mathbf{k}$-dependence as $P(\mathbf{k}) = |a_c(\mathbf{k})|^2$ obtained analytically at the large-field limit $|E_x| \rightarrow \infty$, where no oscillation appears, and scale it by $P_{\mathrm{LZ}}(k_y)$. In Fig.1(b), $P'(\mathbf{k})$ is plotted by the red dashed curves together with $P(\mathbf{k})$, and it reproduces non-oscillatory part of $P(\mathbf{k})$ well. The local transition rate in the incoherent case is given by $dP'(\mathbf{k})/dt = \{\partial P'(\mathbf{k})/\partial k_x\}(dk_x/dt) = W(\mathbf{k})(-eE_x/\hbar)$, where

$$W(\mathbf{k})=\frac{\partial P'(\mathbf{k})}{\partial k_x}=\mathrm{sgn}(-E_x)P_{\mathrm{LZ}}(k_y)\frac{\hbar v_x\left\{\Delta^2+(\hbar v_y k_y)^2\right\}}{2\left\{\Delta^2+(\hbar v_x k_x)^2+(\hbar v_y k_y)^2\right\}^{3/2}}\ . \qquad (11)$$

Since the physical transition rate $dP'(\mathbf{k})/dt$ is always positive, $W(\mathbf{k}) = \partial P'(\mathbf{k})/\partial k_x$ changes sign depending on the sign of $-eE_x/\hbar$, namely the driving direction of $\mathbf{k}$.

As mentioned above, when LZ tunneling occurs in the Dirac system, a Hall shift current can appear as the summation of side jumps. The 2D Hall shift current density along the $y$ direction in the right-handed valley is expressed as

$$j_y^{(\mathrm{shift})}=\frac{2}{(2\pi)^2}\iint_{\mathrm{BZ}}(-e)[\mathbf{R}_{cv}^{(x)}(\mathbf{k})]_y W(\mathbf{k})\left(\frac{-eE_x}{\hbar}\right)f(E_{\mathrm{v}}(\mathbf{k}))\{1-E_{\mathrm{c}}(\mathbf{k})\}dk_x dk_y\ . \qquad (12)$$

The factor 2 corresponds to the spin degeneracy. Here, $f(E)$ denotes the nonequilibrium distribution function under the electric field. The factor $f(E_{\mathrm{v}})\{1-f(E_{\mathrm{c}})\}$ represents Pauli blocking, which inhibits the transition from occupied valence-band states to the occupied conduction-band states. As explained later, the contribution from the deviation $\delta f(E)$ of $f(E)$ from the equilibrium Fermi distribution $f_0(E)$ is cancelled between the two valleys. Therefore, $f(E)$ may be replaced by $f_0(E)$. In a lightly electron-doped system, the factor $f_0(E_{\mathrm{v}})\{1-f_0(E_{\mathrm{c}})\}$ restricts the integral region to the outside of the Fermi surface (FS) at low temperatures. Since the integrand in Eq. (12), except $f(E_{\mathrm{v}})\{1-f(E_{\mathrm{c}})\}$, is an odd function of $\mathbf{k}$, its integral over the Brillouin zone (BZ) vanishes. Therefore, Eq. (12) is equal to −1 times the integral over the inside of the FS. Finally, we obtain the following formula for the 2D Hall shift current density of a single valley at low temperatures:

$$j_y^{(\mathrm{shift})}=\frac{2e^2}{h}|E_x|\frac{1}{2\pi}\iint_{\mathrm{FS}}[\mathbf{R}_{\mathrm{cv}}^{(x)}(\mathbf{k})]_y\,|W(\mathbf{k})|\,dk_x dk_y\ . \qquad (13)$$

The shift vectors $\mathbf{R}_{\mathrm{cv}}^{(x)}(\mathbf{k})$ of the $\mathbf{k}$ states with sufficiently large $|W(\mathbf{k})|$ in the FS

mainly contribute to the total Hall shift current. It depends only on the absolute value of the electric field through the prefactor and $P_{\mathrm{LZ}}(\mathbf{k})$. This means that the sign of the shift current is independent of the electric-field direction and is instead determined by the FS anisotropy and the polarity of the shift vector reflecting the valley chirality. When the direction of the electric field is reversed, the Hall shift current is unchanged; namely, it shows an even-order, rectifying Hall response, as in the conventional NLAHE.

In Fig. 2, the vector field pattern of $\mathbf{R}_{\mathrm{cv}}^{(x)}(\mathbf{k})$ and the magnitude $|W(\mathbf{k})|$ are illustrated by an arrow plot and a density plot, respectively. The FS of the electron-doped tilted Dirac cone is also drawn. When the Dirac cone is tilted, the FS becomes an off-centered ellipse, which is inversion-asymmetric, giving rise to a finite Hall shift current. Figure 2(a) shows the shift vector field around the right-handed valley described by (2), whereas Fig. 2(b) shows that around the time-reversal valley described by (4), where the shift vectors are reversed. In the time-reversal valley, the shift vectors are reversed, and the off-centering of the FS is also reversed. Therefore, the two valleys contribute additively to the Hall shift current; that is, the shift currents from the two valleys have the same direction.

We then examine the effect of the distribution shift $\delta f(E)$ on the total shift current. When $\delta f(E)$ is small under the electric field, $f(E_{\mathrm{c}}(\mathbf{k}))$ is approximately equal to $f_0(E_{\mathrm{c}}(\mathbf{k}+\tau e\mathbf{E}/\hbar))$, where $\tau$ is the scattering relaxation time in the conduction band. As a result, the occupied states are shifted from the equilibrium FS by $\delta\mathbf{k} = -\tau e\mathbf{E}/\hbar$. Since the field-induced distribution shift $\delta\mathbf{k}$ is common to the two valleys, whereas the shift-vector texture is reversed, the Hall shift-current contributions arising from this FS shift cancel between two valleys for small $\delta\mathbf{k}$. This is the reason why we can use $f_0(E)$ instead of $f(E)$ in Eq. (12).

The Hall shift current $j_y^{(\mathrm{shift})}$ for a single valley calculated using Eq. (13) is plotted as a function of the Fermi energy $\mu$ for several electric fields $|E_x|$ at the low-temperature limit in Fig. 3. The Dirac cone tilt is set to $\eta = v_0/v_x = 0.8$. The shift current appears above the band edge $\mu/\Delta = (1-\eta^2)^{1/2} = 0.6$, takes a peak around $\mu/\Delta = 1.5$, and then decreases moderately. As a function of $|E_x|$, $j_y^{\mathrm{(shift)}}/|E_x|$ increases gradually from zero and saturates to a constant value in the large-$|E_x|$ limit, reflecting the field dependence of $P_{\mathrm{LZ}}(k_y)$. This means that $j_y^{\mathrm{(shift)}}$ is proportional to $|E_x|^2$ at small $|E_x|$, but crosses over to a $|E_x|$-linear dependence at the large-$|E_x|$ limit.

We compare this Hall shift current $j_y^{\mathrm{(shift)}}$ with the nonlinear anomalous Hall current $j_y^{\mathrm{(BCD)}}$ arising from the BCD mechanism in the conduction band [15]. The Hall current due to the BCD for a single valley is formulated as follows:

$$j_y^{\mathrm{(BCD)}} = \frac{2e^2}{h}\left|E_x\right| \cdot (-2\pi[\mathbf{\Lambda}_c^z]_x) \cdot \left|[\delta\mathbf{k}]_x\right|, \tag{14}$$

$$[\mathbf{\Lambda}_c^z]_x = \frac{1}{(2\pi)^2}\iint_{\mathrm{FS}} \frac{\partial[\mathbf{\Omega}_\mathrm{c}(\mathbf{k})]_z}{\partial k_x} dk_x dk_y \ , \tag{15}$$

$$[\mathbf{\Omega}_c(\mathbf{k})]_z = -\frac{\hbar v_x \hbar v_y \Delta}{2\left\{\Delta^2 + \left(\hbar v_x k_x\right)^2 + \left(\hbar v_y k_y\right)^2\right\}^{3/2}} . \tag{16}$$

Here, $\mathbf{\Lambda}_\mathrm{c}^z$ is the BCD at low temperatures, and $\mathbf{\Omega}_\mathrm{c}(\mathbf{k})$ is the Berry curvature of the conduction band in the right-handed valley. In Fig. 3, $j_y^{\mathrm{(BCD)}}$ is also plotted by the dashed black curve for $v_x e\tau|E_x|/\Delta = \hbar v_x|[\delta\mathbf{k}]_x|/\Delta = 1$. For general $\tau$ and $|E_x|$ values, $j_y^{\mathrm{(BCD)}}/|E_x|$ is obtained by multiplying this curve by $v_x e\tau|E_x|/\Delta$. As seen in Fig. 3, $j_y^{\mathrm{(BCD)}}$ is of the same order as $j_y^{\mathrm{(shift)}}$ for $v_x e\tau|E_x|/\Delta = 1$, and the $j_y^{\mathrm{(BCD)}}$ curve decays more rapidly at large Fermi energies.

Finally, we estimate $j_y^{\mathrm{(shift)}}$ in comparison with $j_y^{\mathrm{(BCD)}}$ in the weak CO state of $\alpha$-

$(\mathrm{ET})_2\mathrm{I}_3$. From the DFT calculation, the Dirac cone parameters were estimated to be $v_x \sim 10^5$ m/s, $v_y \sim v_x$, and $v_0 = \eta v_x \sim 0.8 v_x$ [16]. In our previous magnetotransport study, we confirmed a gapped Dirac fermion state with the gap parameter $|\Delta| \sim 0.17$ meV in the weak CO state at 1.12 GPa [5]. We have also observed the NLAHE in the weak CO state with a 2D carrier density of $n_{2\mathrm{D}} \sim 1.8\times10^{10}$ cm$^{-2}$, which corresponds to $\mu/\Delta \sim 4.3$ [3]. When we set $\tau \sim 0.2$ ps and $|E_x| \sim 1$ V/mm, which correspond to $\tau/(\hbar/\Delta) \sim 0.05$ and $\hbar v_x e|E_x|/\Delta^2 \sim 2$, the normalized FS shift $\hbar v_x|\delta\mathbf{k}|/\Delta = v_x e\tau|E_x|/\Delta$ is about $0.05 \times 2 = 0.1$, which is sufficiently smaller than the FS size for $\mu/\Delta \sim 4.3$. Since the dotted BCD curve in Fig. 3 is plotted for $v_x e\tau|E_x|/\Delta = 1$, the experimental estimate $v_x e\tau|E_x|/\Delta \sim 0.1$ reduces the BCD contribution by a factor of 0.1. At $\mu/\Delta \sim 4.3$, this gives $j_y^{(\mathrm{BCD})}$ about 1.5 times larger than $j_y^{(\mathrm{shift})}$ for $\hbar v_x e|E_x|/\Delta^2 = 2$, shown by the red-orange curve. Therefore, $j_y^{(\mathrm{shift})}$ is not negligible compared with $j_y^{(\mathrm{BCD})}$ in the weak CO state of $\alpha$-$(\mathrm{ET})_2\mathrm{I}_3$ with an extremely small gap. In actual $\alpha$-$(\mathrm{ET})_2\mathrm{I}_3$ crystals, the observed $j_y^{(\mathrm{shift})}$ is reduced, in the same way as $j_y^{(\mathrm{BCD})}$, by the cancellation between CO domains [15].

In conclusion, we have discussed the NLAHE in 2D gapped Dirac fermion systems. In addition to conventional intraband mechanisms such as the BCD, we propose an interband mechanism arising from LZ tunneling across a small gap. The Hall shift current originates from the side jump associated with LZ tunneling and becomes finite due to FS anisotropy in carrier-doped tilted Dirac cones. Because the Hall shift current depends only on the magnitude of the electric field, it contains only even-order contributions. For an equilibrium FS, the Hall shift currents from the two time-reversal valleys add constructively, whereas the contributions from a nonequilibrium FS shift cancel. The Hall shift current can be comparable in magnitude to the NLAHE arising from the BCD. It provides a possible explanation for the NLAHE accompanied by nonlinear longitudinal transport observed in $\alpha$-$(ET)_2I_3$ with an extremely small gap.

**Acknowledgements**

This work was partially supported by JSPS KAKENHI Grant Numbers JP23K03297, JP24H01610, JP26K06975, and JP26H00589. The author thanks Dr. Andhika Kiswandhi for his experimental comments, and Prof. Tatsuo Hasegawa and Prof. Yasuhiro H. Matsuda for providing an excellent research environment.

**References**

*osada@issp.u-tokyo.ac.jp

**Figure 1** (Osada)

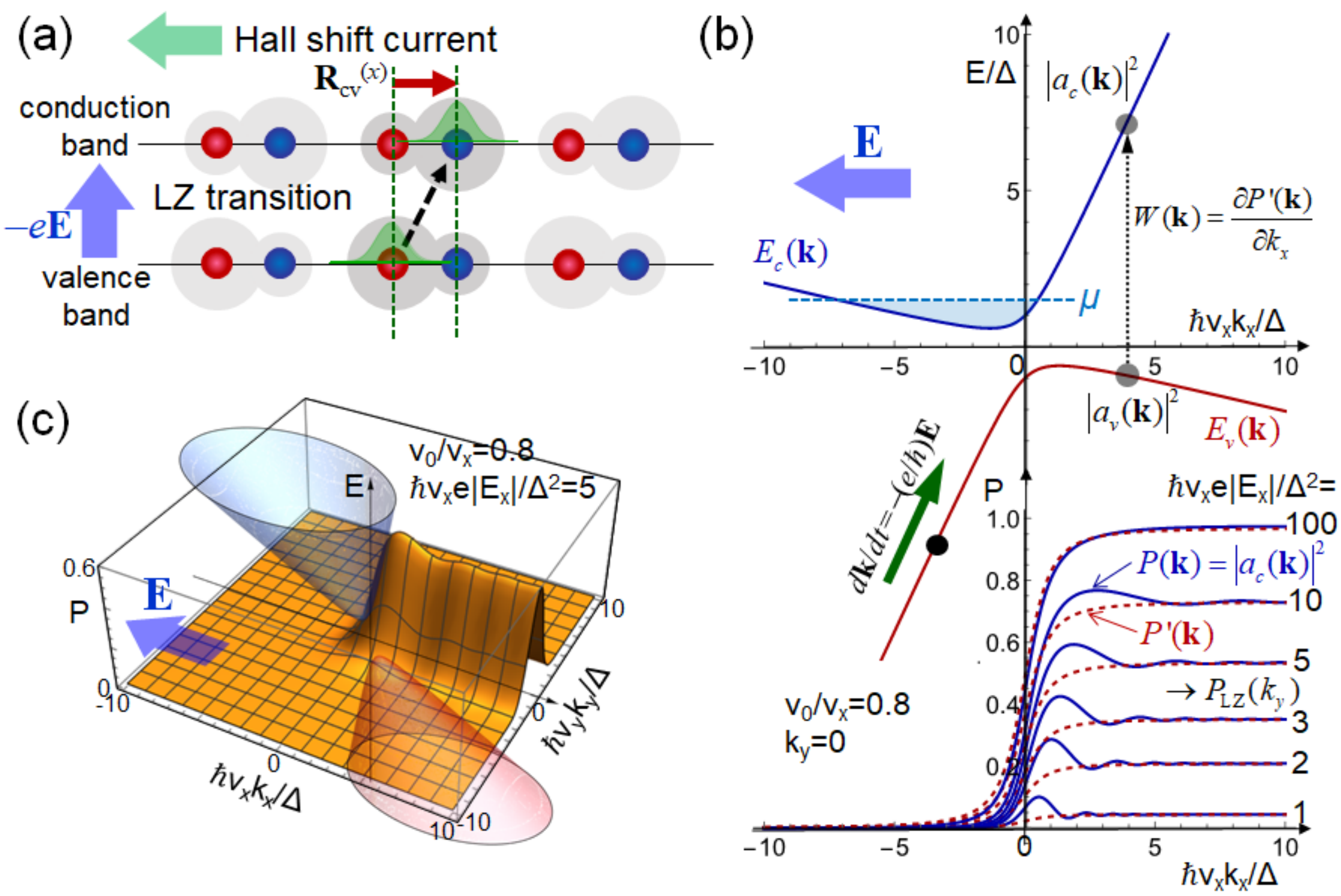

**FIG. 1.** (color online)

(a) Schematic illustration of the shift vector in the LZ transition, representing the shift of the electron-density center associated with LZ tunneling. (b) Schematic illustration of LZ tunneling of a driven electron in a tilted massive Dirac cone (upper panel) and the transient tunneling probability (lower panel). The blue solid and red dashed curves correspond to the coherent and incoherent cases, respectively. (c) Transient tunneling probability in 2D **k** space when an electron is driven along the $k_x$ direction. The tilted gapped Dirac cone is also shown for reference.

**Figure 2** (Osada)

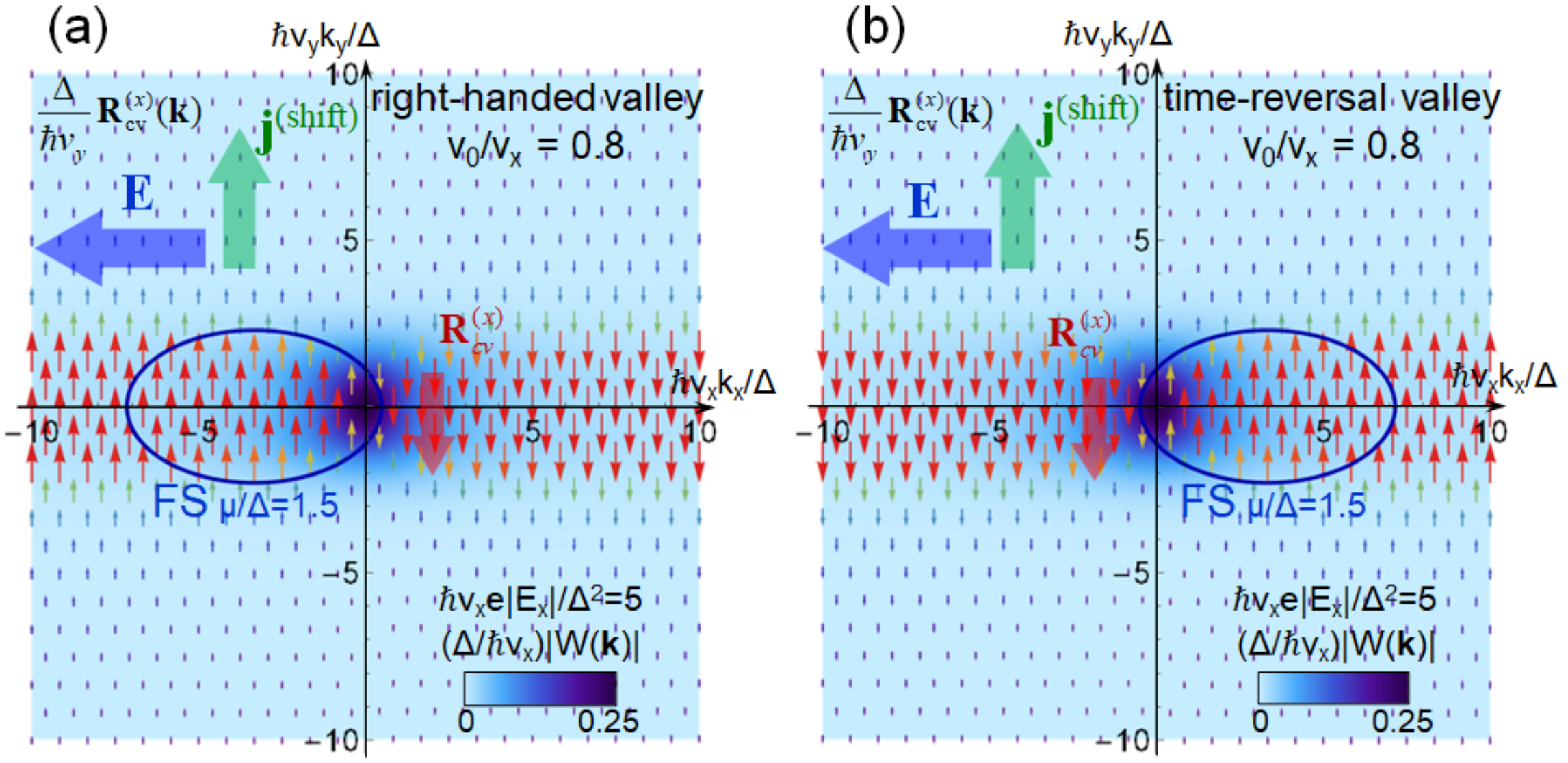

**FIG. 2.** (color online)

(a) Shift-vector field of the 2D right-handed massive Dirac fermion system. The background color represents the magnitude of the local transition rate, $|W(\mathbf{k})|$. The FS of the conduction band in the tilted Dirac cone is also shown. (b) Shift-vector field of the 2D left-handed massive Dirac fermion system, corresponding to the time-reversal partner of (a). The shift vector and the FS are obtained from those in (a) by applying $k_x \rightarrow -k_x$.

**Figure 3** (Osada)

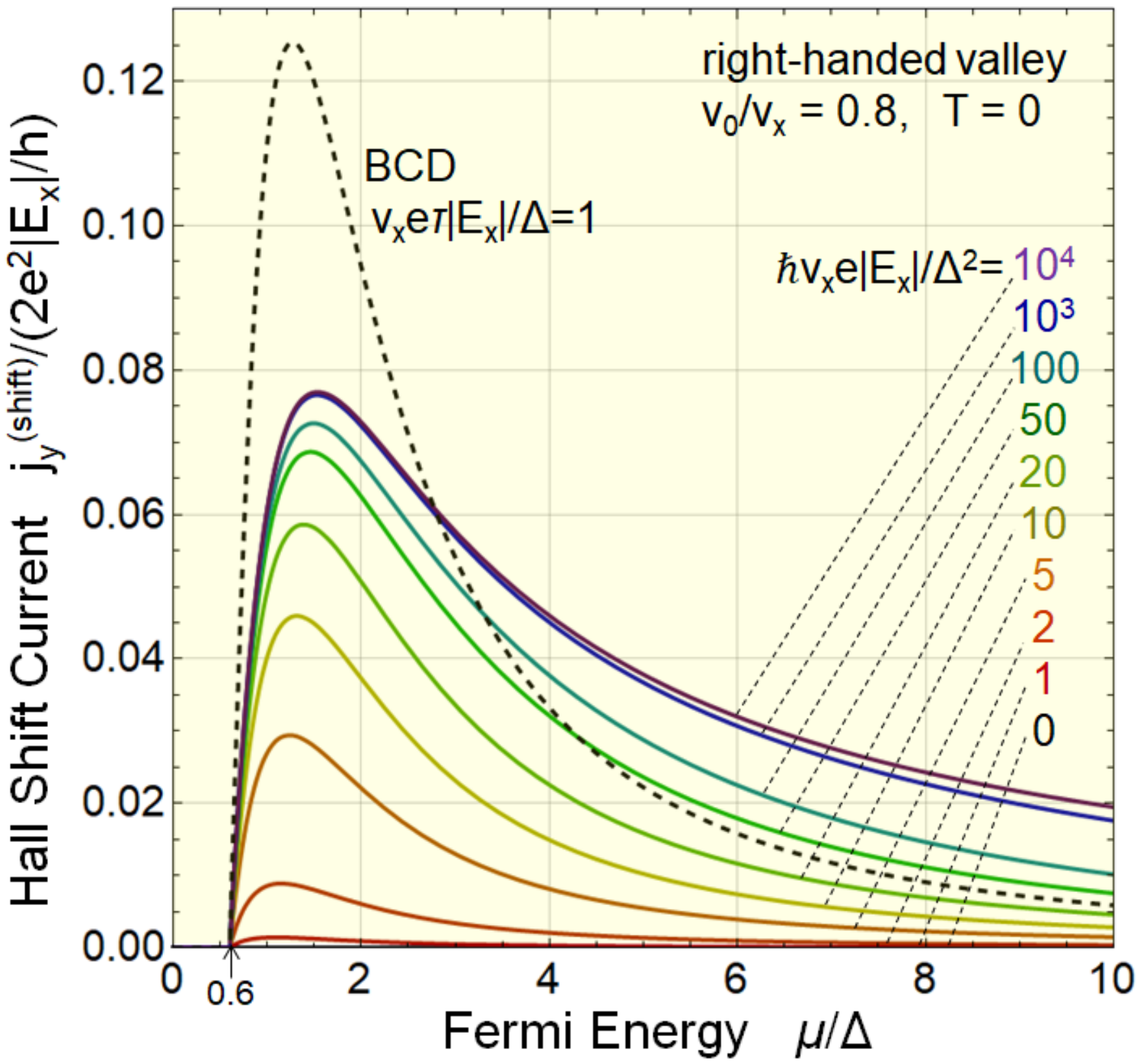

**FIG. 3.** (color online)

Calculated Fermi-energy dependence of the Hall shift current for several electric-field strengths. The Hall shift current is normalized by the quantum Hall current $2e^2|E_x|/h$. The black dashed curve indicates the anomalous Hall current arising from the BCD mechanism in the conduction band for $v_x e\tau|E_x|/\Delta=1$. For given values of $\tau$ and $E_x$, the corresponding BCD Hall current is obtained by multiplying the black dashed curve by $v_x e\tau|E_x|/\Delta$.